\documentclass[aps,twocolumn,showpacs,dvips]{revtex4}
\usepackage{feynmp}
\usepackage{amssymb}
\usepackage{epsfig}
\usepackage{graphicx}
\usepackage{subfigure}
\usepackage{hyperref}
\usepackage{bbm}

\begin{document}

\title{Velocity renormalization of nodal quasiparticles in $d$-wave superconductors}

\author{Jing Wang}
\altaffiliation{jwang315@mail.ustc.edu.cn}
\affiliation{Department of Modern Physics, University of Science and
Technology of China, Hefei, Anhui 230026, P. R. China}

\begin{abstract}
Gapless nodal quasiparticles emerge at a low-energy regime of high-$T_c$ cuprate
superconductors due to the $d_{x^2 - y^2}$ gap symmetry. We study the unusual
renormalizations of the Fermi velocity $v_F$ and gap velocity $v_{\Delta}$ of these
quasiparticles close to various quantum critical points in a superconducting
dome. Special attention is paid to the behavior of the velocity ratio, $v_{\Delta}/v_F$,
since it determines a number of observable quantities. We perform a renormalization-group
analysis and show that the velocity ratio may vanish, approach unity, or diverge at 
different quantum critical points. The corresponding superfluid densities and critical
temperatures are suppressed, slightly increased, or significantly enhanced. The effects
of three types of static disorders, namely, random mass, random gauge potential, and random
chemical potential, on the stability of the system are also addressed. An analogous analysis reveals that both random mass and random gauge potential are irrelevant. This implies that these fixed points of the velocity ratio are stable, and hence observable effects ignited by them
are unchanged. However, the random chemical potential is marginal. As a result, these fixed
points are broken, and thus, the instabilities of quantum phase transitions are triggered.
\end{abstract}


\maketitle


\section{Introduction}

It has been well-established that high-$T_c$ cuprate superconductors
have an anisotropic $d_{x^2 - y^2}$ energy gap. This gap vanishes at
four nodes $\left(\pm \frac{\pi}{2},\pm \frac{\pi}{2}\right)$, so the low-energy
elementary excitations are gapless nodal quasiparticles (QPs) with a linear
energy spectrum. These nodal QPs are responsible for many low-temperature
thermal and transport properties of the superconducting phase of high-$T_c$
superconductors \cite{Orenstein}. For instance, the specific heat exhibits
a linear temperature dependence, $C(T) \propto T$ \cite{Hardy}, which is
quite different from that of conventional $s$-wave superconductors, and
indeed has played a crucial role in the determination of $d_{x^2 - y^2}$
gap symmetry. In the superconducting state, nodal QPs are revealed by
numerous experiments, especially angle-resolved photoemission spectroscopy
(ARPES) \cite{Shen} and transport measurements \cite{Taillefer,Chiao}, to
be well-defined, with extraordinary long mean-free paths \cite{Orenstein}.
A residual short-range interaction between nodal QPs leads only to a quite
weak damping rate, $\propto$ $\max\left(\omega^3, T^3\right)$, and insignificant
corrections to fermion velocities \cite{Hirschfeld}.

In 1999, ARPES measurements by Valla \emph{et al.} \cite{Valla} revealed a
fermion damping rate, $\propto$ $T$, in the optimally doped cuprate superconductor Bi$_{2}$Sr$_{2}$CaCu$_{2}$O$_{8+\delta}$, which indicates a marginal Fermi
liquid behavior \cite{Orenstein00PRL}. This unexpected finding is apparently
in sharp contradiction to theoretical predictions. Such a strong damping can
only be caused by some kinds of soft (gapless) bosonic modes \cite{Hirschfeld}.
In order to explain this experimental finding, Vojta \emph{et al.} \cite{Vojta2000PRB,Vojta2000IJMPB,Vojta2000PRL} proposed that the soft boson
may arise from a quantum phase transition from a pure $d_{x^2- y^2}$ superconducting
state to a new $X$ superconducting state. Generically, there are a number of
candidates for the proposed $X$ state, and it seems difficult to uniquely determine
which is the correct one. Vojta \emph{et al.} \cite{Vojta2000PRB,Vojta2000IJMPB,
Vojta2000PRL} considered all possibilities for the order parameter of the $X$ state,
and ruled out most of them after carrying out careful symmetry analysis and
field-theoretic calculations.

Nodal QPs have two velocities: the Fermi velocity $v_F$ and the gap velocity $v_{\Delta}$ \cite{Durst}. Extensive experiments have determined that they are not equal to each
other, $v_F \neq v_{\Delta}$. Indeed, transport and ARPES measurements \cite{Orenstein,
Chiao} found that the velocity ratio, $v_{\Delta}/v_F \approx 0.1$, in most high-$T_c$
superconductors. Remarkably, a number of important observable quantities depend
on such a velocity ratio \cite{Orenstein}, including the superfluid density \cite{Lee97},
critical temperature $T_c$ \cite{Lee97}, and electric and thermal conductivities \cite{Durst,Mesot,Vojta}. Any unusual renormalization of this velocity ratio will
give rise to considerable changes in these observable quantities. In the presence
of transition from a pure $d$-wave superconducting state to a new $X$ superconducting
state, the fluctuation of the new order parameter can forcefully couple to gapless nodal
QPs near the quantum critical point, which may lead to nontrivial velocity renormalizations.
If this ratio deviates strongly from its bare value, these physical quantities will be either enhanced or suppressed.

In this paper, we focus on the unusual renormalizations of fermion velocities
caused by the critical fluctuations of different $X$ order parameters in the
$d$-wave superconductor. We are particularly interested in the low-energy
behavior of the velocity ratio $v_{\Delta}/v_F$. Inspired by the study of velocity renormalization in graphene \cite{Son2007,WangjrLiu,WangLiu2}, it can be analyzed by
the renormalization-group (RG) method \cite{Huh, Shankar,WLK}. The unusual ratio
$v_{\Delta}/v_F$ will have significant impacts on transport properties. Moreover,
the transport properties of nodal QPs are largely determined by scattering due
to impurities. It is, therefore, necessary to examine the influence of various
impurity potentials, other than the interaction between nodal QPs and $X$ ordering,
on the behavior of fermion velocities. Based on the coupling between nodal QPs and
disorders, there are three types of disorders in a $d$-wave superconductor: random mass,
random gauge potential, and random chemical potential \cite{Stauber}. We include these
disordered potentials in our model and explore their effects by means of the RG method.

Clean-limit systems are considered first. We employ three types of vertex
matrices, $M=\tau^x$, $\tau^y$, and $\tau^z$, to denote various quantum phase transitions.
[Quantum phase transitions from a pure $d_{x^2- y^2}$ to a new $X$ superconducting
state can be classified by vertex matrices, $M$, between nodal QPs and $X$ order
parameters \cite{Vojta2000PRL}, shown in  Eq. (\ref{Eq_S_psi-phi}), and henceforth
we dub the new $X$ superconducting state the $X$ matrix state.] By implementing
the RG analysis \cite{Huh, Shankar,WLK}, we obtain three distinct fixed points of the
velocity ratio due to the fluctuations between nodal QPs and $X$ order parameters in
the vicinity of the quantum critical points, which are $v_\Delta/v_F\rightarrow0$ \cite{Huh},
$v_\Delta/v_F\rightarrow1$, and $v_\Delta/v_F\rightarrow\infty$ for $M=\tau^x$, $\tau^y$,
and $\tau^z$, respectively. The case $v_\Delta/v_F\rightarrow0$ has been discussed
in recent publications \cite{Huh,WLK}, the others will be focused on in this paper.
Since many physical properties display $v_\Delta/v_F$ dependence, observable effects
kindled by these interesting fixed points should be expected. Indeed, we find that
superfluid density and critical temperature are sensitively influenced by approaching
these fixed points. Both are suppressed, slightly increased and significantly
enhanced for cases $M=\tau^x$, $\tau^y$, and $\tau^z$, respectively.

Whether the results in the clean limit are stable against the disorders are also investigated.
In practice, the disorder effects induced by various kinds of scattering are inevitably present
in the low-temperature transport properties of an interacting electron system. If the fixed points are changed or even broken by the disorder effects, the behavior of physical observables will be considerably affected. Hence, it is imperative to examine the disorder effects on RG flows of fermion velocities. In general, there are three types of disorders (random mass,
random gauge potential, and random chemical potential \cite{Stauber}) coupled to gapless nodal
QPs in the $d$-wave superconductor. The impacts of these disorders on the low-temperature transport properties of nodal QPs have been studied extensively \cite{Nersesyan,Altland}. In our case, the RG flows of fermion velocities will be influenced by these disordered potentials, and in the meanwhile, the RG flows of strength parameters of fermion-disorder couplings are determined by fermion velocities. Therefore, we should self-consistently compute the flows of fermion velocities with disorder strength parameters. After a detailed RG analysis of the interplay between $X$ order parameter fluctuations and disorder scattering, a series of coupled
RG equations of Fermi velocity $v_F$, gap velocity $v_\Delta$, and disorder strength parameter $v_\Gamma$ are derived. Based on numerical calculations, we learn that both random mass and random gauge potential are irrelevant. This signifies that these two types of disorders do not change the flows of fermion velocities. Accordingly, the corresponding fixed points and physical observables are stable. However, the random chemical potential is marginal. Consequently, these fixed points will be destroyed and hence the instabilities of quantum phase transitions are signaled.

The rest of the paper is organized as follows. The effective field theory
and the corresponding Feynman rules are presented in Sec. \ref{Sec_model}.
We calculate the self-energy and vertex corrections in Sec. \ref{Sec_self-energy}
and Sec. \ref{Sec_vertex}, respectively. A detailed RG analysis is given in
Sec. \ref{Sec_RG-analysis}, which is followed by discussions of the numerical
solutions of the RG equations in Sec. \ref{Sec_numerical} and of the behaviors of
the superfluid density and critical temperature caused by velocity renormalization
in Sec. \ref{sec_quantities}. Finally, we briefly summarize our results in Sec. \ref{Sec_summary}.

\section{Effective field theory of quantum critical phenomena}\label{Sec_model}

We begin with the action
\begin{eqnarray}
S &=& S_{\Psi} + S_{\phi} + S_{\Psi\phi},
\end{eqnarray}
where the free action for nodal QPs is
\begin{eqnarray}\!\!\!\!\!\!\!\!\!\!\!\!\!\!\!\!\!\!\!
S_{\Psi} \!&=&\!\!\! \int\frac{d^{2}\mathbf{k}}{(2\pi)^{2}}\frac{d\omega}{2\pi}
\Psi^{\dagger}_{1a}(-i\omega+v_{F}k_{x}\tau^{z} +
v_{\Delta}k_{y}\tau^x)\Psi_{1a} \nonumber \\
&&\hspace{-1.4em} +\int\frac{d^{2}\mathbf{k}}{(2\pi)^2}\frac{d\omega}{2\pi}
\Psi^{\dagger}_{2a}(-i\omega+v_{F}k_{y}\tau^{z} +
v_{\Delta}k_{x}\tau^{x})\Psi_{2a},\label{2}
\end{eqnarray}
where $\tau^{(x,y,z)}$ denote Pauli matrices. The linear dispersion of Dirac fermions
originates from the $d_{x^2 - y^2}$-wave symmetry of the energy gap of cuprate
superconductor. Here, the Nambu spinor $\Psi^{\dagger}_{1}$ represents nodal QPs
excited from the $(\frac{\pi}{2},\frac{\pi}{2})$ and $(-\frac{\pi}{2},-\frac{\pi}{2})$
nodal points, and $\Psi^{\dagger}_{2}$ represents the other two nodal points; $\omega$ is a Matsubara frequency in the zero-temperature limit, $k_{x,y}$ describe the wave vector from
the nodal points and have been rotated by $\frac{\pi}{2}$ and $v_F$ and $v_\Delta$ are the
Fermi velocity and the gap velocity, respectively \cite{Huh}. The repeated spin index $a$ is summed from 1 to $N_f$, the number of fermion spin components. The ratio $v_{\Delta}/v_{F} \approx 1/20$ between the Fermi velocity and the gap velocity is determined by experiments \cite{Orenstein, Chiao}.

With the help of group-theoretic classification, Vojta \emph{et al.} \cite{Vojta2000PRB,Vojta2000IJMPB,Vojta2000PRL} pointed out seven
possible quantum phase transitions from a pure $d_{x^2-y^2}$ superconducting
state to $X$ matrix states. The effective action $S_{\phi}$, which
describes the $X$ order parameter in real space is
\begin{eqnarray}
S_{\phi} = \int d^2\mathbf{x}d\tau\Big\{\frac{1}{2}(\partial\tau
\phi)^2 + \frac{c^2}{2}(\nabla\phi)^2 +
\frac{r}{2}\phi^2+\frac{u_0}{24}\phi^4\Big\},
\end{eqnarray}
where $\tau$ is imaginary time and $c$ is the velocity. The mass parameter $r$
drives the system to undergo quantum phase transitions, with $r=0$ defining
the zero-temperature quantum critical point. $u_0$ is the quartic self-interaction
strength. The interaction between nodal QPs $\Psi_{1,2}$ and the $X$ order parameter
$\phi$ is described by a Yukawa coupling \cite{Vojta2000PRL,Huh}
\begin{eqnarray}
S_{\Psi\phi} = \int d^2\mathbf{x}d \tau
\left[\lambda_0\phi(\Psi^\dagger_1M_1\Psi_1
+\Psi^\dagger_2M_2\Psi_2)\right],\label{Eq_S_psi-phi}
\end{eqnarray}
with $\lambda_0$ the coupling constant. There are seven possible $X$ matrix states,
which can be distinguished by different choices of matrices $M_1$ and $M_2$ \cite{Vojta2000PRL,Vojta2000IJMPB}. It was shown in Refs.~\cite{Vojta2000PRL,
Vojta2000IJMPB} that two states out of these seven candidates are irrelevant.
Therefore, we only need to concentrate on the remainder of the states: (i) $M_1 = \tau^y$,
$M_2 = \tau^y$; (ii) $M_1 =\tau^y$, $M_2 = -\tau^y$; (iii)$M_1 = \tau^x$,
$M_2 = \tau^x$; (iv) $M_1= \tau^z$, $M_2 = -\tau^z$; and (v)$M_1 = \tau^x$,
$M_2 = -\tau^x$.

The state with $M_1 = M_2 = \tau^x$ corresponds to a nematic state, which has
been extensively investigated in a number of papers \cite{Kivelson,Kivelson03,
Fradkin,Ando,Hinkov,Daou,Lawler,Mackenzie,Castellani,Kim,Huh, Xu08,Sachdev09,
Fritz09,WLK,Liu,WangLiu,Metzner,Oganesyan,Metzner1,Metzner2,Metzner3,Rech,Sunkai,
Garst,Metlitski}. The critical fluctuation of such nematic order leads to an
extreme velocity anisotropy and other unusual properties. The influence of
disorders was studied in Ref.~\cite{WLK}. In this paper, we consider the other
four states. Fortunately, when calculating the fermion self-energy and boson
polarization, $M_{1}$ or $M_{2}$ always appears in pairs. Hence the contributions
do not depend on the signs of Pauli matrices. For instance, cases i and ii or
iii and v would lead to the same results. In order to simplify the discussion,
we can ignore the possible minus of $M_2$ to obtain a compact classification.
Therefore, we have only two possibilities: $M = \tau^y$ and $M = \tau^z$. (The
basic conclusions are independent of this simplification.)

\begin{figure}
\includegraphics[width=2.1in]{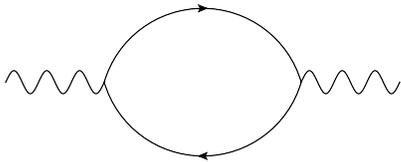}
\caption{The polarization function for the $X$ order parameter. The
solid line represents the fermion propagator and the wavy line
represents the boson propagator.}\label{Fig_polarization}
\end{figure}

To analyze the above field theory, we consider a general and large
fermion flavor $N_f$, and apply a $1/N_f$ expansion. The free
fermion propagator is
\begin{eqnarray}
G^{0}_{\Psi}(\mathbf{k},\omega) = \frac{1}{-i\omega +
v_{F}k_{x}\tau^{z} + v_{\Delta}k_{y}\tau^{x}}.\label{Eq_fermion-propagator}
\end{eqnarray}
for nodal QPs $\Psi_1$ (the free propagator for nodal QPs $\Psi_2$ can be
written similarly).

To the leading order of $1/N_f$ expansion, the polarization function
is shown in Fig. (\ref{Fig_polarization}) and symbolizes the integral
\begin{eqnarray}
\Pi(\mathbf{q},\epsilon) =
\int\frac{d^{2}\mathbf{k}}{(2\pi)^{2}}\frac{d\omega}{2\pi}
\mathrm{Tr}[M G^{0}_{\Psi}(\mathbf{k},\omega)M
G^{0}_{\Psi}(\mathbf{k+q},\omega+\epsilon)], \nonumber
\end{eqnarray}
After straightforward calculations \cite{Huh,WLK}, we have
\begin{eqnarray}
\Pi(\mathbf{q},\epsilon) &\equiv& \Pi^y(\mathbf{q},\epsilon)
\nonumber \\
&=& \frac{1}{16v_{F}v_{\Delta}}\sqrt{\epsilon^2 +v_{F}^{2}q_{x}^{2}
+ v_{\Delta}^{2}q_{y}^{2}} \nonumber \\
&& +\frac{1}{16v_{F}v_{\Delta}}\sqrt{\epsilon^{2}+v_{F}^{2}q_{y}^{2}
+ v_{\Delta}^{2}q_{x}^{2}}
\end{eqnarray}
for the case of $M = \tau^y$, and
\begin{eqnarray}
\Pi(\mathbf{q},\epsilon) &\equiv& \Pi^z(\mathbf{q},\epsilon)
\nonumber \\
&=& \frac{1}{16v_{F}v_{\Delta}} \frac{(\epsilon^{2} +
v_{\Delta}^{2}q_{y}^{2})}{(\epsilon^2 +
v_{F}^{2}q_{x}^{2}+v_{\Delta}^{2}q_{y}^{2})^{1/2}} \nonumber \\
&& + \frac{1}{16v_{F}v_{\Delta}}\frac{(\epsilon^{2} + v_{\Delta}^{2}
q_{x}^{2})}{(\epsilon^{2}+v_{F}^{2}q_{y}^{2} +
v_{\Delta}^{2}q_{x}^{2})^{1/2}}
\end{eqnarray}
for $M = \tau^z$. In a low energy regime, the polarization function is
linear in $|q|$ and, therefore dominates over the $q^2$-term. Near
the quantum critical point, we keep only the mass term and assume
that $\phi \longrightarrow \phi/\lambda_0$ and $r\longrightarrow N_f
r \lambda^2_0$, leading to \cite{Huh}
\begin{eqnarray}
S = S_{\Psi} \!+\!\! \int\! \!d^2\mathbf{x} d \tau
\Big\{\frac{N_fr}{2}\phi^2 \!+ \phi[\Psi^{\dagger}_{1a} M
\Psi_{1a}\! + \Psi^{\dagger}_{2a}M \Psi_{2a}]\Big\}\!.\label{9}
\end{eqnarray}
After integrating out the fermion degrees of freedom, the effective
action for the scalar field ($X$ order parameter) becomes
\begin{eqnarray}\label{7}
\frac{S_{\phi}}{N_f} = \frac{1}{2}\int \frac{d^3q}{(2\pi)^3}
[r+\Pi(q)]|\phi(q)|^{2} +\mathcal{O}(\phi^{4}).\label{10}
\end{eqnarray}
Now the effective propagator of the $X$ order parameter is
\begin{eqnarray}
G_{\phi}(\mathbf{q},\epsilon)=\frac{1}{\Pi(\mathbf{q},
\epsilon)}\label{Eq_nematic-propagator}
\end{eqnarray}
at the quantum critical point $r = 0$. Propagators (\ref{Eq_fermion-propagator})
and (\ref{Eq_nematic-propagator}) will be utilized in the following calculations of
the fermion self-energy and RG equations.

\begin{figure}
\includegraphics[width=3.4in]{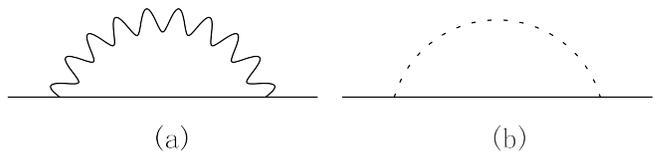}
\caption{One-loop fermion self-energy correction due to (a) $X$-order-parameter
fluctuation and (b) disorder. The dashed line represents disorder scattering.}\label{Fig_self-energy}
\end{figure}

In almost all realistic condensed matter systems, disorders are present
and play significant roles in determining the low-temperature behaviors.
In the present problem, the nodal QPs can interact with three
types of random potentials, which represent different disorder scattering
processes. According to the coupling between nodal QPs and disorders,
there are three kinds of random potentials in $d$-wave superconductors: random mass,
random chemical potential, and random gauge potential. All these types of
disorders have been investigated in the contexts of the $d$-wave cuprate superconductor
\cite{Nersesyan, Altland}, quantum Hall effect \cite{Ludwig}, and graphene
\cite{Stauber, Aleiner}. In the subsequent analysis, we consider
all three sorts of disorders.

The coupling term between the fermion field and a random field $A(\mathbf{x})$
can be written as \cite{Stauber}
\begin{eqnarray}
\int d^2 \mathbf{x}
\Psi^{\dagger}(\mathbf{x})\Gamma\psi(\mathbf{x})A(\mathbf{x}).
\end{eqnarray}
The matrixes $\Gamma$ are $\Gamma =\mathrm{I}$, $\Gamma = \tau^{y}$, and $\Gamma = (\tau^{x},\tau^{z})$ for a random chemical potential, a random mass, and a random
gauge potential, respectively. We assume that the random potential $A(\mathbf{x})$ is a
quenched, Gaussian white-noise field with the correlation functions
\begin{eqnarray}
\langle A(\mathbf{x})\rangle = 0; \hspace{0.5cm} \langle
A(\mathbf{x}_1)A(\mathbf{x}_2)\rangle = g
v_{\Gamma}^{2}\delta^2(\mathbf{x}_1 - \mathbf{x}_2),
\end{eqnarray}
where the dimensionless parameter $g$ represents the concentration of
impurity, and the parameter $v_\Gamma$ measures the strength of a
single impurity. It will be convenient to redefine the random
potential as $A(\mathbf{x})\rightarrow v_{\Gamma}A(\mathbf{x})$,
and then write the fermion-disorder interaction term as \cite{Stauber}
\begin{eqnarray}
S_{\mathrm{dis}} = v_{\Gamma}\int d^2 \mathbf{x}
\Psi^{\dagger}(\mathbf{x})\Gamma\psi(\mathbf{x})A(\mathbf{x}),
\end{eqnarray}
with the random potential distribution
\begin{eqnarray}\label{A_equation}
\langle A(\mathbf{x})\rangle = 0; \hspace{0.5cm} \langle
A(\mathbf{x}_1)A(\mathbf{x}_2)\rangle = g\delta^2(\mathbf{x}_1 -
\mathbf{x}_2).
\end{eqnarray}
Then by studying the vertex correction to the fermion-disorder
interaction term, we can obtain the RG flow of disorder strength.
After a Fourier transformation, the corresponding action for the
fermion-disorder interaction has the form
\begin{eqnarray}
S_{\mathrm{dis}} = v_{\Gamma}\!\!\int \!d^{2}\mathbf{k}
d^{2}\mathbf{k}_1d\omega \Psi^{\dagger}(\mathbf{k},\omega)\Gamma
\Psi(\mathbf{k}_1,\omega)A(\mathbf{k-k_1}).
\end{eqnarray}
This action is analyzed together with the actions (\ref{2}),
(\ref{9}), and (\ref{10}). In order to carry out perturbative expansion,
both $g$ and $v_\Gamma$ are assumed to be small in magnitude, corresponding
to the weak disorder case.

Before making RG analysis of the velocities and disorder strength parameter,
we calculate the one-loop fermion self-energy and vertex corrections in the
following two sections.

\section{Fermion self-energy corrections}\label{Sec_self-energy}

The interplay between $X$-order-parameter fluctuation and random
potentials can yield self-energy corrections to the free propagator
of nodal QPs, which are described by the Dyson equation
\begin{eqnarray}
G^{-1}_{\Psi}(\mathbf{k},\omega) &=& -i\omega + v_F k_x \tau^z +
v_{\Delta} k_{y}\tau^{x} \nonumber \\
&& -\Sigma_{\mathrm{X}}(\mathbf{k},\omega) -
\Sigma_{\mathrm{dis}}(\mathbf{k},\omega),
\end{eqnarray}
where self-energy functions $\Sigma_{\mathrm{X}}(\mathbf{k},\omega)$
and $\Sigma_{\mathrm{dis}}(\mathbf{k},\omega)$ come from X order parameter
fluctuation and disorder scattering, respectively. To the leading order,
the corresponding Feynman diagrams of self-energy are presented in Fig.
(\ref{Fig_self-energy}).

Employing the method of Ref. \cite{Huh}, we obtain
\begin{eqnarray}
\frac{d\Sigma^{y}_{\mathrm{X}}(\mathbf{k},\omega)}{d\ln\Lambda} =
C_1(-i\omega) + C_2 v_F k_x \tau^z + C_3 v_{\Delta} k_y \tau^x,
\end{eqnarray}
where
\begin{widetext}
\begin{eqnarray}
C_1&=&\frac{2(v_\Delta/v_F)}{N_f \pi^3}\int^{\infty}_{-\infty} dx
\int^{2\pi}_{0} d \theta \frac{x^2-\cos^2\theta-(v_\Delta/v_F)^2
\sin^2\theta}{(x^2+\cos^2\theta+(v_\Delta/v_F)^2
\sin^2\theta)^2}\mathcal {G}(x,\theta),\label{Eq_C_1}\\
C_2&=&\frac{2(v_\Delta/v_F)}
{N_f \pi^3}\int^{\infty}_{-\infty}\!\!\!
dx \int^{2\pi}_{0} \!\!d \theta
\frac{-x^2+\cos^2\theta-(v_\Delta/v_F)^2
\sin^2\theta}{(x^2+\cos^2\theta+(v_\Delta/v_F)^2
\sin^2\theta)^2}\mathcal {G}(x,\theta),\label{Eq_C_2}\\
C_3&=&\frac{2(v_\Delta/v_F)}
{N_f \pi^3}\int^{\infty}_{-\infty}\!\!\!
dx \int^{2\pi}_{0} \!\!d \theta\frac{-x^2
-\cos^2\theta+(v_\Delta/v_F)^2 \sin^2\theta }
{(x^2+\cos^2\theta+(v_\Delta/v_F)^2
\sin^2\theta)^2}\mathcal {G}(x,\theta),\label{Eq_C_3}\\
\mathcal{G}^{-1}&=&\sqrt{x^2+\cos^2\theta+(v_\Delta/v_F)^2 \sin^2\theta}
+ \sqrt{x^2 +\sin^2\theta+(v_\Delta/v_F)^2\cos^2\theta},\label{Eq_mathcal_G}
\end{eqnarray}
for $M=\tau^y$, and
\begin{eqnarray}
\frac{d\Sigma^{z}_{\mathrm{X}}(\mathbf{k},\omega)}{d\ln\Lambda} =
D_1(-i\omega) + D_2 v_F k_x \tau^z + D_3 v_{\Delta} k_y \tau^x,
\end{eqnarray}
where
\begin{eqnarray}
D_1&=&\frac{2(v_F/v_\Delta)}{N_f \pi^3}\int^{\infty}_{-\infty} dx
\int^{2\pi}_{0} d \theta \frac{x^2-\cos^2\theta-(v_\Delta/v_F)^2
\sin^2\theta}{(x^2+\cos^2\theta+(v_\Delta/v_F)^2
\sin^2\theta)^2}\mathcal {F}(x,\theta),\label{Eq_D_1}\\
D_2&=&\frac{2(v_F/v_\Delta)}
{N_f \pi^3}\int^{\infty}_{-\infty}\!\!\!
dx \int^{2\pi}_{0} \!\!d \theta
\frac{x^2-\cos^2\theta+(v_\Delta/v_F)^2
\sin^2\theta}{(x^2+\cos^2\theta+(v_\Delta/v_F)^2
\sin^2\theta)^2}\mathcal {F}(x,\theta),\label{Eq_D_2}\\
D_3&=&\frac{2(v_F/v_\Delta)}
{N_f \pi^3}\int^{\infty}_{-\infty}\!\!\!
dx \int^{2\pi}_{0} \!\!d \theta\frac{-x^2
-\cos^2\theta+(v_\Delta/v_F)^2 \sin^2\theta }
{(x^2+\cos^2\theta+(v_\Delta/v_F)^2
\sin^2\theta)^2}\mathcal {F}(x,\theta),\label{Eq_D_3}\\
\mathcal{F}^{-1} &=& \frac{x^2+\sin^2\theta}{\sqrt
{x^2+\cos^2\theta+(v_\Delta/v_F)^2 \sin^2\theta}} +
\frac{x^2+\cos^2\theta}{\sqrt{x^2 +\sin^2\theta
+(v_\Delta/v_F)^2\cos^2\theta}} \label{Eq_mathcal_F}
\end{eqnarray}
for $M = \tau^z$.
\end{widetext}

The fermion self-energy due to disorder
$\Sigma_{\mathrm{dis}}(i\omega)$ can be computed as
\begin{eqnarray}
\Sigma_{\mathrm{dis}}(i\omega) &=& gv^2_\Gamma
\int\frac{d^{2}\mathbf{k}}{(2\pi)^2}\Gamma
G^{0}_{\psi}(\mathbf{k},\omega)\Gamma \nonumber \\
&=& \frac{g v_{\Gamma}^{2}}{2\pi v_{F}v_{\Delta}}i\omega\ln\Lambda.
\end{eqnarray}
According to this result, we infer that $\Sigma_{\mathrm{dis}}(i\omega)$
exhibits the same behavior for all possible choices of $\Gamma$ and
is actually $M$-matrix independent. Another striking feature is
that $\Sigma_{\mathrm{dis}}(i\omega)$ does not depend on momentum,
which reflects the fact that quenched disorders are static. This
leads to
\begin{eqnarray}
\frac{d\Sigma_{\mathrm{dis}}(i\omega)}{d\ln\Lambda} = C_g i\omega,
\end{eqnarray}
where
\begin{eqnarray}
C_g = \frac{g v_{\Gamma}^{2}}{2\pi v_{F} v_{\Delta}}.
\end{eqnarray}

\section{Vertex corrections}\label{Sec_vertex}

\begin{figure}
\includegraphics[width=3.35in]{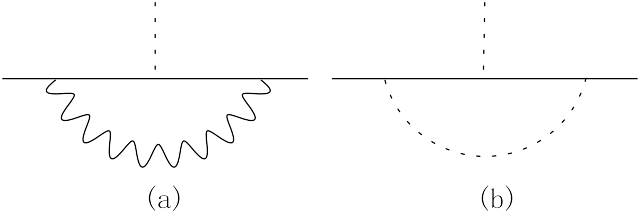}
\caption{Fermion-disorder vertex correction due to (a) the $X$ order
parameter and (b) the disorder parameter.}\label{Fig_vertex}
\end{figure}

The fermion-disorder interaction parameter $v_\Gamma$ is also subjected
to RG flow. To obtain its flow equation, we need to calculate the
fermion-disorder vertex corrections. Formally, the vertex correction
has the form
\begin{equation}
v_{\Gamma}\Gamma' = v_{\Gamma}\Gamma + \Gamma_{i} +V_{i},
\end{equation}
where $\Gamma_{i}$ represents the vertex correction due to
$X$ order parameter fluctuation and $V_{i}$ represents the vertex
correction due to disorder interaction, and the quantities denoted
$i=1,2,3$ correspond to the random chemical potential, random mass
and random gauge potential, respectively. The Feynman diagrams are
shown in Fig. (\ref{Fig_vertex}). Both cases, $M = \tau^y$ and $M = \tau^z$,
are calculated explicitly in the following for all three types
of disorders.

\subsection{Random chemical potential}

To compute the vertex correction owing to $X$ ordering, we take advantage
of the method proposed by Huh and Sachdev \cite{Huh}. At zero external
momenta and frequencies, the vertex correction is expressed as
\begin{eqnarray}
\Gamma_{1} = v_{\Gamma} \int\frac{d^{3}Q}{(2\pi)^3}H(Q)
\mathcal{K}^{3}\left(\frac{\mathbf{q}^{2}}{\Lambda^{2}}\right),
\end{eqnarray}
where $\mathcal {K}(x)$ is an arbitrary function with $\mathcal
{K}(0)=1$, and it falls off rapidly with $x$, e.g., $\mathcal
{K}(x)=e^{-x}$. However, the results are independent of the
particular choices of $\mathcal {K}(x)$. There is a useful
formula \cite{Huh},
\begin{eqnarray}
\frac{d\Gamma_{1}}{d\ln\Lambda} = v_{\Gamma}
\frac{v_{F}}{8\pi^3}\int^{\infty}_{-\infty}dx \int ^{2\pi}_{0}
d\theta H(\hat{Q}),
\end{eqnarray}
where
\begin{eqnarray}
H(\hat{Q}) &=& \frac{1}{N_f}\tau^i\frac{1}{(-iv_{F}x +
v_{F}\cos\theta\tau^{z} + v_{\Delta}\sin\theta\tau^x)}\mathrm{I}
\nonumber \\
&& \times \frac{1}{(-iv_{F}x + v_{F}\cos\theta\tau^{z} +
v_{\Delta}\sin\theta\tau^{x})}\tau^{i}\frac{1}{\Pi(\hat{Q})}.
\nonumber \\
\end{eqnarray}
where $i=y$ and $z$ denote types $M=\tau^y$ and $M=\tau^z$, respectively.
Here, matrix $\mathrm{I}$ corresponds to the coupling between
nodal QPs and the random chemical potential. It will be replaced by
$\tau^{y}$ in the case of random mass and $\tau^{x,z}$ in the case
of random gauge potential. After straightforward calculation, we have
\begin{eqnarray}
\frac{d\Gamma_{1}}{d\ln\Lambda}
&=& \left\{\begin{array}{ll}
C_{5}v_{\Gamma}\mathrm{I},\hspace{0.5cm} M=\tau^y
\\\\
D_{5}v_{\Gamma}\mathrm{I}, \hspace{0.5cm}M=\tau^z.
\end{array}\right.
\end{eqnarray}
where
\begin{eqnarray}
C_{5} &=& -\frac{2(v_\Delta/v_F)}{N_f\pi^3}\int^\infty_{-\infty} dx
\int^{2\pi}_{0} d\theta \nonumber \\
&& \times \frac{(x^2 - \cos^{2}\theta -
(v_{\Delta}/v_{F})^{2}\sin^{2}\theta)} {(x^{2} + \cos^{2}\theta +
(v_{\Delta}/v_{F})^{2}\sin^{2}\theta)^{2}}\mathcal{G}(x,\theta)
\nonumber \\
&=& -C_1,
\end{eqnarray}
and
\begin{eqnarray}
D_{5} &=& -\frac{2(v_F/v_\Delta)}{N_f\pi^3}\int^\infty_{-\infty} dx
\int^{2\pi}_{0} d\theta \nonumber \\
&& \times \frac{(x^2 - \cos^{2}\theta -
(v_{\Delta}/v_{F})^{2}\sin^{2}\theta)} {(x^{2} + \cos^{2}\theta +
(v_{\Delta}/v_{F})^{2}\sin^{2}\theta)^{2}}\mathcal{F}(x,\theta)
\nonumber \\
&=& -D_1.
\end{eqnarray}

The vertex correction due to averaging over disorder is
\begin{eqnarray}
V_{1} = gv_{\Gamma}^{2}
\int\frac{d^{2}\mathbf{p}}{(2\pi)^{2}} \mathrm{I}
G^0_{\Psi}(\omega,\mathbf{p})v_{\Gamma}\mathrm{I}
G^0_{\Psi}(\omega,\mathbf{p+k})\mathrm{I}.
\end{eqnarray}
Again, one should substitute a certain Pauli matrix for matrix $\mathrm{I}$
in the case of random mass or random gauge potential. Taking the external momentum
$\mathbf{k}=0$ and keeping only the leading divergent term, we have
\begin{eqnarray}
\frac{dV_{1}}{d\ln\Lambda} =
C_{\Gamma}v_{\Gamma}\mathrm{I},
\end{eqnarray}
where
\begin{eqnarray}
C_{\Gamma} = \frac{v_{\Gamma}^{2} g}{2\pi v_{F}v_{\Delta}} = C_{g},
\end{eqnarray}
for both types.

\subsection{Random mass}

Calculation of the vertex correction in the case of random mass
parallels the process presented above, so we just state the final
result. The $X$-ordering-induced vertex correction is
\begin{eqnarray}
\frac{d\Gamma_{2}}{d\ln\Lambda}
&=& \left\{\begin{array}{ll}
C_{6}v_{\Gamma}\tau^{y},\hspace{0.5cm} M=\tau^y
\\\\
D_{6}v_{\Gamma}\tau^{y}, \hspace{0.5cm}M=\tau^z.
\end{array}\right.
\end{eqnarray}
where
\begin{eqnarray}
C_{6} &=& \frac{2(v_\Delta/v_F)}{N_f\pi^3}\int^\infty_{-\infty} dx
\int ^{2\pi}_0d \theta \nonumber \\
&& \times \frac{(-x^2-\cos^2\theta -
(v_{\Delta}/v_{F})^2\sin^2\theta)}{(x^2+\cos^2\theta +
(v_{\Delta}/v_{F})^2\sin^2\theta)^2}\mathcal{G}(x,\theta) \nonumber \\
&=& C_{3}-C_{1}-C_{2},
\end{eqnarray}
\begin{eqnarray}
D_{6} &=& \frac{2(v_F/v_\Delta)}{N_f\pi^3}\int^\infty_{-\infty} dx
\int ^{2\pi}_0d \theta \nonumber \\
&& \times \frac{(x^2+\cos^2\theta +
(v_{\Delta}/v_{F})^2\sin^2\theta)}{(x^2+\cos^2\theta +
(v_{\Delta}/v_{F})^2\sin^2\theta)^2}\mathcal{F}(x,\theta)\nonumber \\
&=& D_{2}-D_{1}-D_{3}.
\end{eqnarray}

The disorder-induced vertex correction, in both cases, can be written as
\begin{eqnarray}
\frac{dV_{2}}{d\ln\Lambda} =
-C_{\Gamma}(v_{\Gamma}\tau^{y}),
\end{eqnarray}
where
\begin{eqnarray}
C_{\Gamma} = \frac{v_{\Gamma}^{2} g}{2\pi v_{F}v_{\Delta}} = C_{g}.
\end{eqnarray}

\subsection{Random gauge potential}

The random gauge potential has two components, characterized by
$\tau^x$ and $\tau^z$ respectively. For the $\tau^x$ component,
the $X$-ordering contribution to vertex correction is
\begin{eqnarray}
\frac{d\Gamma_{3}}{d\ln\Lambda}
&=& \left\{\begin{array}{ll}
C_{4A}v_{\Gamma}\tau^{x},\hspace{0.5cm} M=\tau^y
\\\\
D_{4A}v_{\Gamma}\tau^{x}, \hspace{0.5cm}M=\tau^z.
\end{array}\right.
\end{eqnarray}
where
\begin{eqnarray}
C_{4A} &=& \frac{2(v_{\Delta}/v_{F})}{N_f\pi^3}
\int^\infty_{-\infty} dx \int^{2\pi}_{0}d\theta \nonumber \\
&& \times \frac{(x^{2}+\cos^{2}\theta -
(v_{\Delta}/v_{F})^{2}\sin^{2}\theta)} {(x^{2} + \cos^{2}\theta +
(v_{\Delta}/v_{F})^{2}\sin^{2}\theta)^{2}} \mathcal{G}(x,\theta)
\nonumber \\
&=& -C_{3},
\end{eqnarray}
\begin{eqnarray}
D_{4A} &=& \frac{2(v_{F}/v_{\Delta})}{N_f\pi^3}
\int^\infty_{-\infty} dx \int^{2\pi}_{0}d\theta \nonumber \\
&& \times \frac{(x^{2}+\cos^{2}\theta -
(v_{\Delta}/v_{F})^{2}\sin^{2}\theta)} {(x^{2} + \cos^{2}\theta +
(v_{\Delta}/v_{F})^{2}\sin^{2}\theta)^{2}} \mathcal{F}(x,\theta)
\nonumber \\
&=& -D_{3}.
\end{eqnarray}

For the $\tau^z$ component, we have
\begin{eqnarray}
\frac{d\Gamma_{3}}{d\ln\Lambda}
&=& \left\{\begin{array}{ll}
C_{4B}v_{\Gamma}\tau^{z},\hspace{0.5cm} M=\tau^y
\\\\
D_{4B}v_{\Gamma}\tau^{z}, \hspace{0.5cm}M=\tau^z.
\end{array}\right.
\end{eqnarray}
where
\begin{eqnarray}
C_{4B} &=& \frac{2(v_{\Delta}/v_{F})}{N_{f}\pi^{3}}
\int^{\infty}_{-\infty}dx\int^{2\pi}_{0}d\theta \nonumber \\
&& \times \frac{(x^{2} - \cos^{2}\theta +
(v_{\Delta}/v_{F})^{2}\sin^{2}\theta)} {(x^{2} + \cos^{2}\theta +
(v_{\Delta}/v_{F})^{2}\sin^{2}\theta)^{2}} \mathcal{G}(x,\theta)
\nonumber \\
&=& -C_{2},
\end{eqnarray}
\begin{eqnarray}
D_{4B} &=&\frac{2(v_{F}/v_{\Delta})}{N_{f}\pi^{3}}
\int^{\infty}_{-\infty}dx\int^{2\pi}_{0}d\theta \nonumber \\
&& \times \frac{(-x^{2} + \cos^{2}\theta -
(v_{\Delta}/v_{F})^{2}\sin^{2}\theta)} {(x^{2} + \cos^{2}\theta +
(v_{\Delta}/v_{F})^{2}\sin^{2}\theta)^{2}} \mathcal{F}(x,\theta)
\nonumber \\
&=& -D_{2}.
\end{eqnarray}

The disorder contribution can be calculated similarly. For both the
$\tau^x$ and the $\tau^z$ components, we have
\begin{eqnarray}
V_{3}(\omega) = \mathrm{finite},
\end{eqnarray}
for types $M=\tau^y$ and $M=\tau^z$. So
\begin{eqnarray}
\frac{dV_{3}}{d\ln\Lambda} = 0.
\end{eqnarray}

\section{RG analysis}\label{Sec_RG-analysis}

In this section, we make an RG analysis of the fermion velocities and
disorder strength and then derive RG equations. To this end, it is convenient
to perform the scaling transformations \cite{Huh, Shankar,WLK}
\begin{eqnarray}
k &=& k'e^{-l}, \\
\omega &=& \omega'e^{-l}, \\
\Psi_{1,2}(\mathbf{k},\omega)
&=& \Psi'_{1,2}(\mathbf{k'},\omega')e^{\frac{1}{2}
\int^{l}_{0}(4 - \eta_{f})dl}, \\
\phi(\mathbf{k},\omega) &=&
\phi'(\mathbf{k'},\omega')e^{\frac{1}{2}\int^{l}_{0} (5 -
\eta_{b})dl},
\end{eqnarray}
where $b = e^{-l}$ with $l>0$. The parameters $\eta_f$ and $\eta_b$
are determined by the self-energy and $X$-ordering-fermion vertex
corrections. Note that the energy is required to rescale in the same way as
the momentum, so the fermion velocities are forced to flow under RG
transformations.

The standard procedure for assigning the scaling transformation of a field
operator when the energy and momentum are rescaled, according to the spirit
of RG theory \cite{Shankar}, is to keep its kinetic term invariant. Since
the random potential $A(\mathbf{x})$ does not possess its own kinetic term,
it actually does not work in the present problem. In order to find out its
scaling behavior, we write the Gaussian white-noise distribution in the
momentum space as
\begin{eqnarray}\label{Gauss_distribution}
\langle A(\mathbf{k}_1)A(\mathbf{k}_2) \rangle =
g\delta^2(\mathbf{k}_1 + \mathbf{k}_2).
\end{eqnarray}
When the momentum $\mathbf{k}$ becomes $b \mathbf{k}$, the delta
function is rescaled to
\begin{eqnarray}
\delta^2(\mathbf{k}_1 + \mathbf{k}_2)\rightarrow
\delta^2(b\mathbf{k}_1 + b\mathbf{k}_2) = b^{-2}
\delta^2(\mathbf{k}_1 + \mathbf{k}_2).
\end{eqnarray}
If we require that the disorder distribution, Eq. (\ref{Gauss_distribution}),
is invariant under scaling transformations, then the random potential should
be transformed as
\begin{eqnarray}
A(\mathbf{k}) \rightarrow b^{-1} A(\mathbf{k}).
\end{eqnarray}
Now we have to assume that
\begin{eqnarray}
A(\mathbf{k}) = A'(\mathbf{k'})e^{l}.
\end{eqnarray}

In the light of the RG technique introduced in Refs. \cite{Son2007}, \cite{WangLiu2},
and \cite{Shankar}, the momentum shell between $b\Lambda$ and $\Lambda$ will be
integrated out, while keeping the $-i\omega$ term invariant. From the message of
type $M=\tau^y$ ordering and disorder contributions to the fermion self-energy
function, we have
\begin{eqnarray}
&& \int^{b\Lambda}d^{2}\mathbf{k} d\omega \Psi^{\dagger}
\!\left[-i\omega -C_1(-i\omega)\ln\frac{\Lambda}{b\Lambda} +
C_g(-i\omega)\ln\frac{\Lambda}{b\Lambda}\right]\!\Psi \nonumber \\
&& = \int^{b\Lambda}d^{2}\mathbf{k} d\omega \Psi^{\dagger}
(-i\omega)[1+(C_g - C_1)l]\Psi \nonumber \\
&& \approx \int^{b\Lambda}d^{2}\mathbf{k} d\omega
\Psi^{\dagger}(-i\omega)e^{(C_g - C_1)l}\Psi.\label{Eq_RG_eta}
\end{eqnarray}
After the scaling transformation, this term should go back to the
free form, so that
\begin{eqnarray}
\eta_{f} = C_{g} - C_{1}.\label{Eq_eta}
\end{eqnarray}
The kinetic terms should also remain invariant under scaling
transformations, which leads to
\begin{eqnarray}
\frac{dv_{F}}{dl} &=& (C_{1}-C_{2}-C_{g})v_{F},\label{Eq_RG_vF} \\
\frac{dv_{\Delta}}{dl} &=& (C_{1}-C_{3}-C_{g})v_{\Delta}.\label{Eq_RG_vD}
\end{eqnarray}
Based on these expressions, the ratio between the gap velocity and the Fermi
velocity is given by
\begin{eqnarray}\label{Eq_RG_vDF}
\frac{d(v_{\Delta}/v_{F})}{dl} = (C_{2}-C_{3})(v_{\Delta}/v_{F}).
\end{eqnarray}
By replacing $C$ with $D$ in Eqs. (\ref{Eq_RG_eta}, \ref{Eq_eta},
\ref{Eq_RG_vF}, \ref{Eq_RG_vD}, \ref{Eq_RG_vDF}), we could get
similar equations for the case $M=\tau^z$.

The disorder strength parameter $v_\Gamma$ enters the above expressions.
Because of the interplay of $X$ ordering and disorder, this parameter also
runs under RG transformations. The flow equation depends on the type
of disorder, which is studied in the following.

We first consider the case of the random chemical potential for $M=\tau^y$.
The bare fermion-disorder action is
\begin{eqnarray}
v_{\Gamma}\int d^{2}\mathbf{k}  d^{2}\mathbf{k}_1d\omega
\Psi^{\dagger}(\mathbf{k},\omega)\Gamma
\Psi(\mathbf{k}_1,\omega)A(\mathbf{k-k_1}).
\end{eqnarray}
Taking into account corrections due to ordering and disorder interactions
yields
\begin{eqnarray}
&&\int^{b\Lambda} d^{2}\mathbf{k}
d^{2}\mathbf{k}_1d\omega\Psi^{\dagger}(\mathbf{k},\omega)
\left[v_{\Gamma}\mathrm{I}- C_{1}v_{\Gamma}\mathrm{I}
\ln\frac{\Lambda}{b\Lambda}\right.\nonumber \\
&& + \left.C_{g}v_{\Gamma}\mathrm{I}\ln\frac{\Lambda}{b\Lambda}\right]
\Psi(\mathbf{k}_1,\omega)A(\mathbf{k-k_1}) \nonumber \\
&=& \int^{b\Lambda}d^{2}\mathbf{k}d^{2}\mathbf{k}_1d\omega
\Psi^{\dagger}(\mathbf{k},\omega)v_{\Gamma}\mathrm{I}
[1+(C_{g}-C_{1})l]\nonumber \\
&& \times \Psi(\mathbf{k}_1,\omega) A(\mathbf{k-k_1}) \nonumber \\
&\approx& \int^{b\Lambda}d^{2}\mathbf{k}d^{2}\mathbf{k}_1d\omega
\Psi^{\dagger}(\mathbf{k},\omega)v_{\Gamma}\mathrm{I}
e^{(C_{g}-C_{1})l}\nonumber \\
&& \times \Psi(\mathbf{k}_1,\omega) A(\mathbf{k-k_1}).
\end{eqnarray}
After redefining the energy, momentum, and field operators, we are left with
\begin{eqnarray}
&& \int^{\Lambda}d^{2}\mathbf{k'}d^{2}\mathbf{k'}_1 d\omega'
\Psi'^{\dagger}(\mathbf{k'},\omega')v_{\Gamma}\mathrm{I} \nonumber \\
&& \times e^{(C_{g}-C_{1})l} \Psi'(\mathbf{k'}_1,\omega')
e^{-\eta_{f}l} A'(\mathbf{k}'-\mathbf{k}'_1). \label{Eq_vGamma}
\end{eqnarray}
Since $\eta_{f} = C_{g} - C_{1}$, it is easy to obtain the following
RG flow equation for $v_{\Gamma}$,
\begin{eqnarray}
\frac{dv_{\Gamma}}{dl}=0\label{Eq_RG_vGamma}
\end{eqnarray}
Evidently, the parameter $v_{\Gamma}$ does not flow and thus can be
simply taken to be a constant. By applying the similar steps, we could
get the same result for $M=\tau^z$.

\begin{figure}[t]
\centering
\epsfig{file=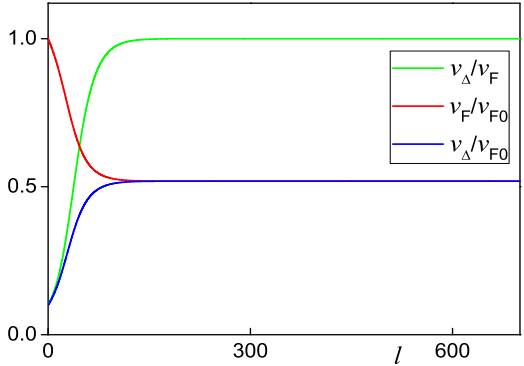,height = 6.15cm,width=9.2cm}\\ 
\vspace{-0.8cm}
\caption{(Color online) Flows of $v_F,v_\Delta,$ and $v_\Delta/v_F$ at a representative initial
value $v_{\Delta0}/v_{F0}=0.1$ for $M=\tau^y$. The conclusions are independent
of these concrete values.}\label{Fig_clean_y}
\end{figure}

In the case of random mass, the running equations for fermion velocities
have the same forms as Eq. (\ref{Eq_RG_vF}) and Eq. (\ref{Eq_RG_vD}).
However, the flow equation for disorder strength parameter is different f
rom Eq. (\ref{Eq_vGamma}), and would be recast into
\begin{eqnarray}
\frac{dv_{\Gamma}}{dl}
&=& \left\{\begin{array}{ll}
(C_{3} - C_{2} - 2C_{g})v_{\Gamma},\hspace{0.5cm} M=\tau^y,
\\\\
(D_{2} - D_{3} - 2D_{g})v_{\Gamma}, \hspace{0.5cm}M=\tau^z,
\end{array}\right.
\end{eqnarray}
which couples self-consistently to the flow equations of the fermion
velocities.

Following the steps presented above, we can derive the corresponding
RG equations in the case of random gauge potential for $M=\tau^y$
\begin{eqnarray}
\frac{d v_F}{dl}&=&(C_1 - C_2 - C_{gi})v_F,
\label{Eq_RG_vFg}\\
\frac{d v_\Delta}{dl}&=&(C_1 - C_3 - C_{gi})v_\Delta,
\label{Eq_RG_vDg}
\end{eqnarray}
which couple to the flow equations of disorder strength
\begin{eqnarray}
\frac{dv_{\Gamma1}}{dl}&=&[(C_1-C_{g1})-C_3]v_{\Gamma1},
\label{Eq_RG_vGg1} \\
\frac{dv_{\Gamma2}}{dl}&=&[(C_1-C_{g2})-C_2]v_{\Gamma2},
\label{Eq_RG_vGg2}
\end{eqnarray}
where
\begin{eqnarray}
C_{g_i} = \frac{v_{\Gamma i}^2 g}{2\pi v_F
v_\Delta},\hspace{0.2cm}i=1,2.\label{Eq__C_gauge}
\end{eqnarray}
Here, the equations denoted by $i=1,2$ correspond to the $\tau^x$
and $\tau^z$ components of the random gauge potential, respectively.
Their $M=\tau^z$-case counterparts are conveniently obtained by
substituting $D$ for $C$ in Eqs. (\ref{Eq_RG_vFg}, \ref{Eq_RG_vDg},
\ref{Eq_RG_vGg1}, \ref{Eq_RG_vGg2}, \ref{Eq__C_gauge}).

\section{Numerical results}\label{Sec_numerical}

In this section, the numerical solutions of the RG equations are presented
and the physical implications of these results are also discussed. We first
consider the clean limit and then include random potentials.

\subsection{Clean limit}\label{Subsec_clean}

By analyzing the coupled RG equations of fermion velocities $v_F$ and $v_\Delta$
introduced in Sec. \ref{Sec_RG-analysis} with clean limit $g=0$, we can obtain
the running flows with decreasing energy scale, i.e., growing scale of $l$.  An extreme
anisotropy of fermion velocities, $(v_\Delta/v_F)^*=0$, caused by the nematic order
parameter ($M=\tau^x$), was found in Ref. \cite{Huh}. We subsequently
list the results of two other cases.

In the case of $M = \tau^y$, the velocity ratio $v_{\Delta}/v_{F}$ flows to a new
fixed point, $(v_\Delta/v_F)^* = 1$, at the lowest energy, as shown in
Fig. (\ref{Fig_clean_y}). This implies that the system becomes isotropic
in the low-energy regime.

\begin{figure}[t]
\centering
\epsfig{file=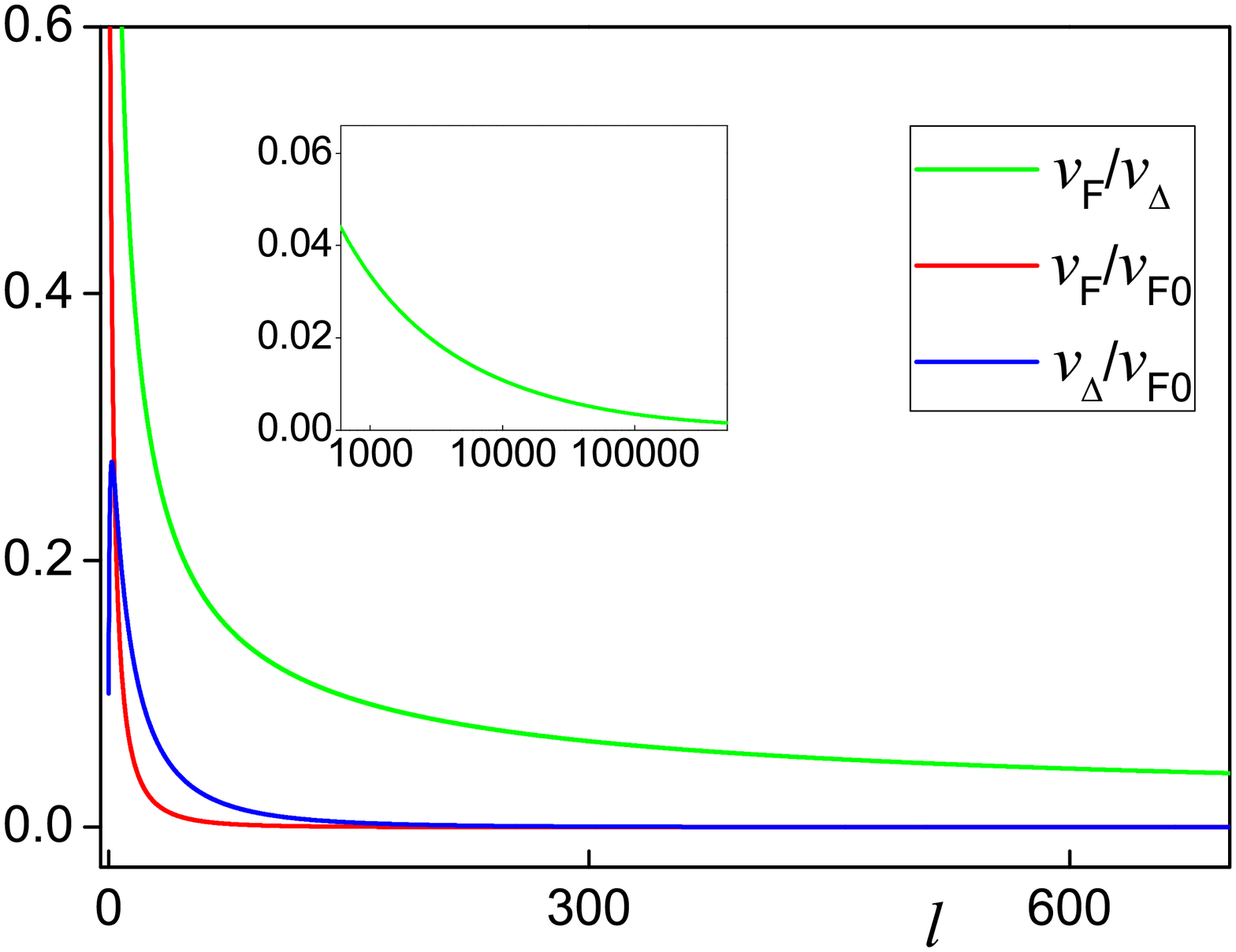,height = 6.15cm,width=9.2cm}\\ 
\vspace{-0.8cm}
\caption{(Color online) Flows of $v_F,v_\Delta,$ and $v_\Delta/v_F$ at the representative
initial value $v_{\Delta0}/v_{F0}=0.1$ for $M=\tau^z$. The conclusions are independent
of these concrete values. Inset: Flow of $v_F/v_\Delta$ for large $l$ ($\geq600$).}\label{Fig_clean_z}
\end{figure}

Regarding $M = \tau^z$, we find another extreme fixed point with $v_{F}/v_{\Delta}
\rightarrow 0$ in the low-energy regime, as depicted in Fig. (\ref{Fig_clean_z}).
(The ratio $v_{F}/v_{\Delta}$ decreases rather slowly as $l$ increases, and this
is discussed in \ref{Subsec_disorder}.) This indicates that the $X$-ordering quantum
phase transition in the $d$-wave superconductors is accompanied by the appearance
of an infinite velocity anisotropy. It is interesting to compare this extreme
anisotropy with that of $v_{\Delta}/v_{F} \rightarrow 0$ driven by the critical
nematic fluctuation \cite{Huh}.

\subsection{Including disorder}\label{Subsec_disorder}

The conclusions in the previous subsection are valid for clean systems. In fact,
disorders are present in almost all realistic condensed matter systems and
play important roles in determining the low-temperature behaviors. In the
current problem, the nodal QPs can interact with three types of disordered
potentials as presented in Sec. \ref{Sec_model}, which represent different
disorder scattering processes. In the general analysis that follows, we
consider the influence of all these kinds of disorders.

\begin{figure}[t]
\centering
\epsfig{file=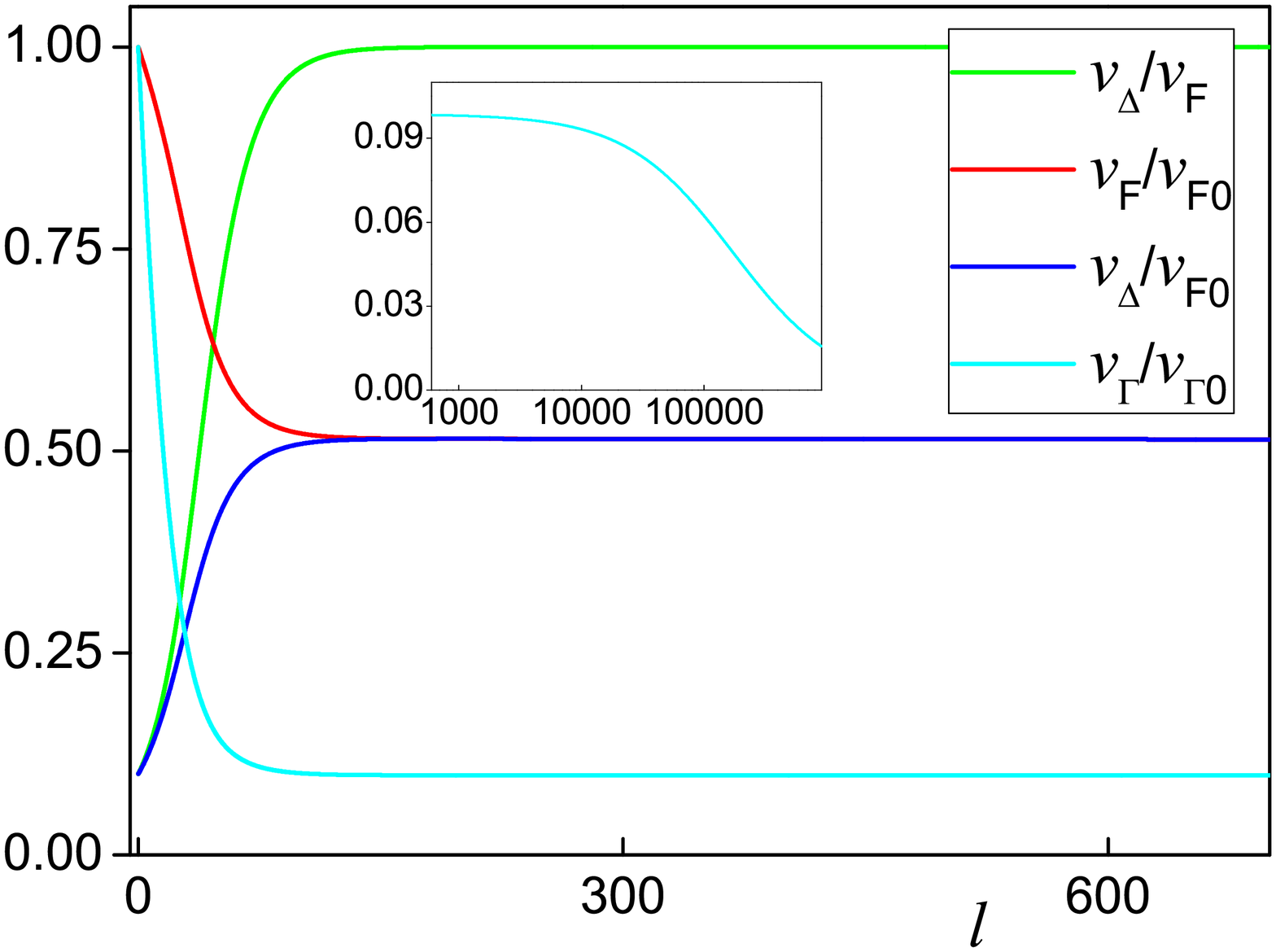,height = 6.15cm,width=9.2cm}\\ 
\vspace{-0.8cm}
\caption{(Color online) Flows of $v_F,v_\Delta,v_\Delta/v_F,$ and $v_\Gamma$ in the presence of
random mass at two representative initial values, $v_{\Delta0}/v_{F0}=0.1$, and $g=10^{-3}$,
for $M=\tau^y$. The conclusions are independent of these concrete values. Inset: Flow of $v_\Gamma$ for large $l$ ($\geq600$).} \label{Fig_mass_y}
\end{figure}

\begin{figure}[t]
\centering
\epsfig{file=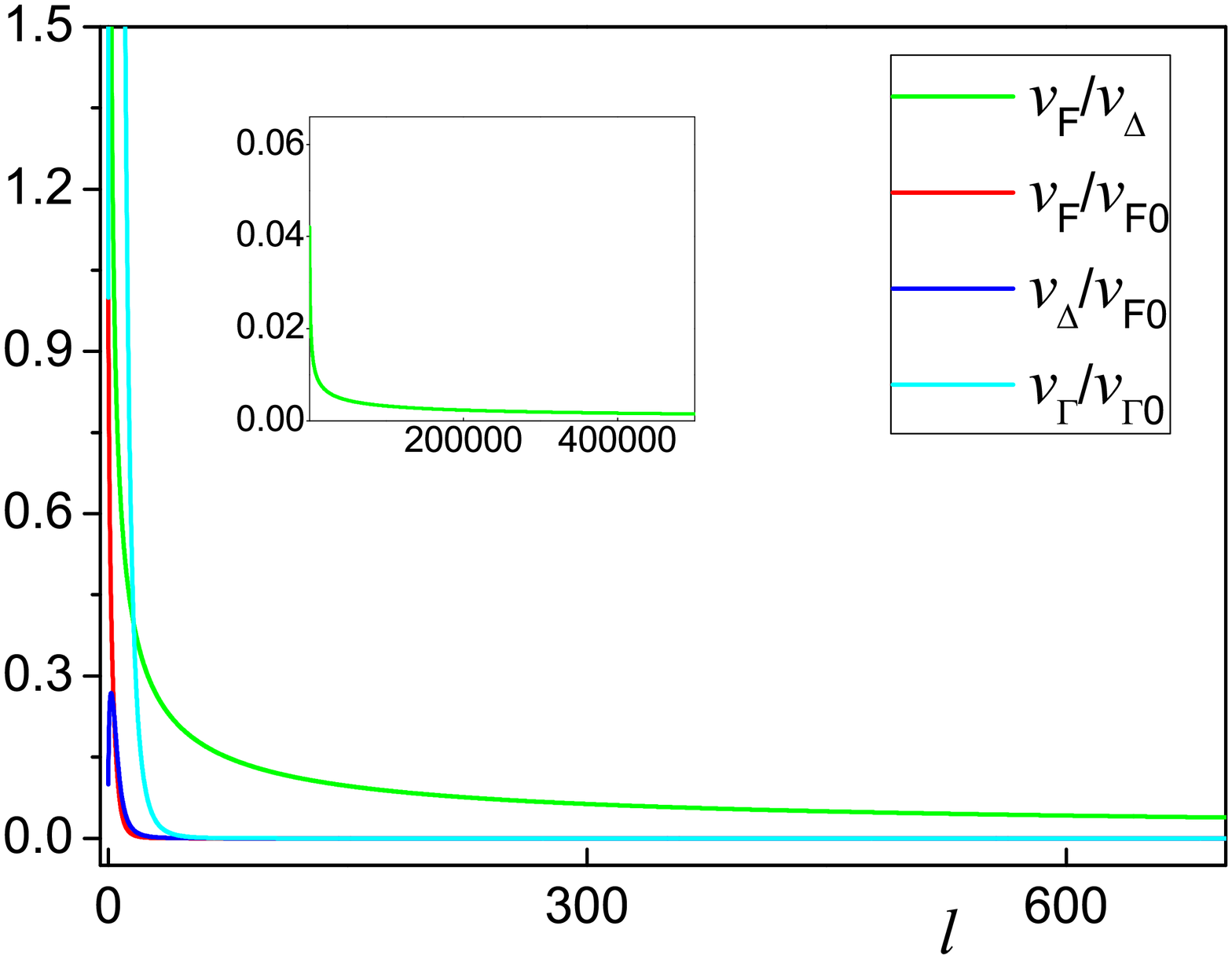,height = 6.15cm,width=9.2cm}
\vspace{-0.8cm}
\caption{(Color online) Flows of $v_F,v_\Delta,v_\Delta/v_F,$ and $v_\Gamma$ in the presence of
random mass at two representative initial values, $v_{\Delta0}/v_{F0}=0.1$, and $g=10^{-3}$, for
$M=\tau^z$. The conclusions are independent of these concrete values. Inset: Flow of $v_F/v_\Delta$ for large $l$ ($\geq600$).} \label{Fig_mass_z}
\end{figure}

\begin{figure}[t]
\centering
\epsfig{file=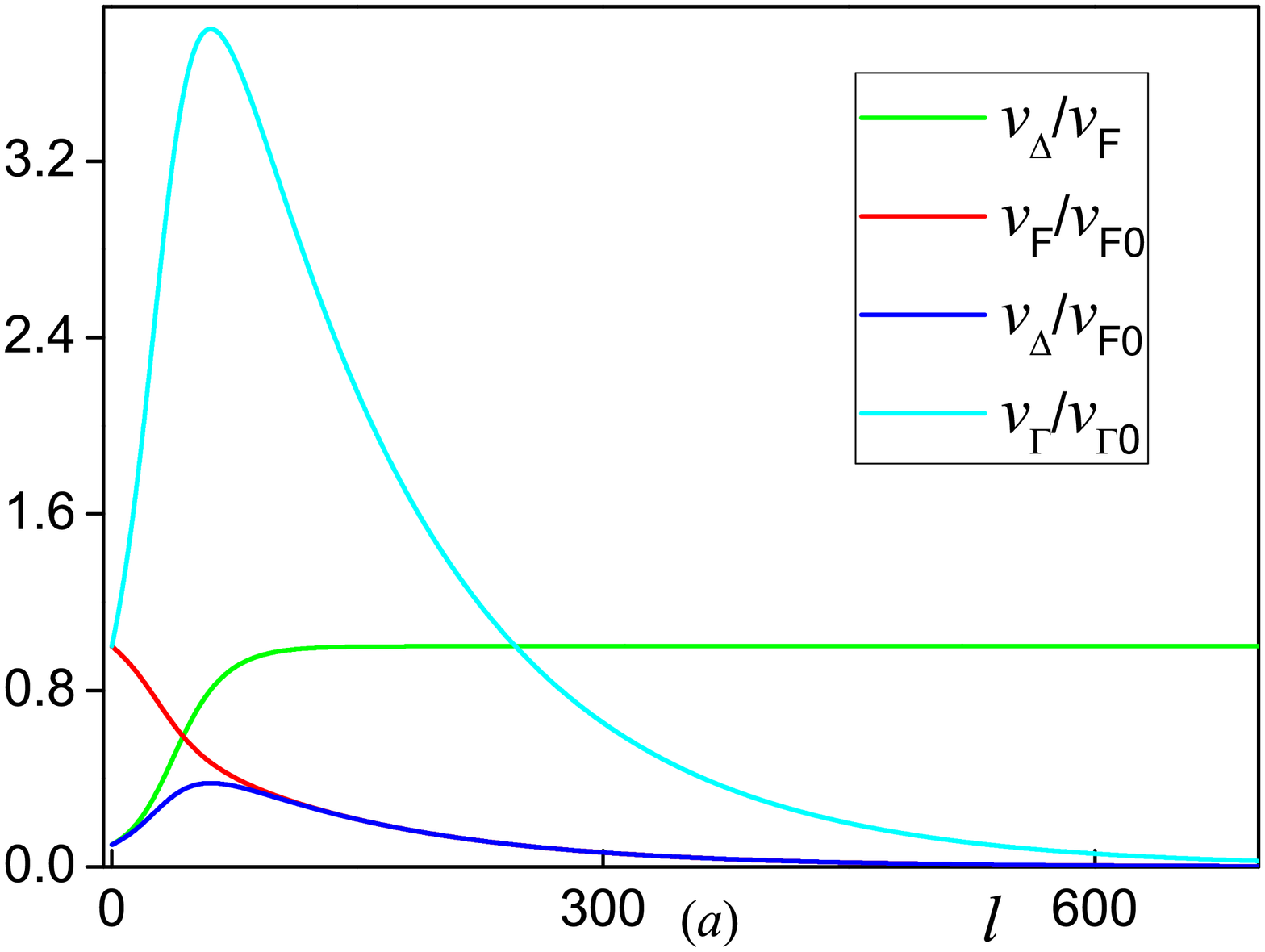,height = 6.15cm,width=9.2cm}\\ \vspace{-1.0cm}
\epsfig{file=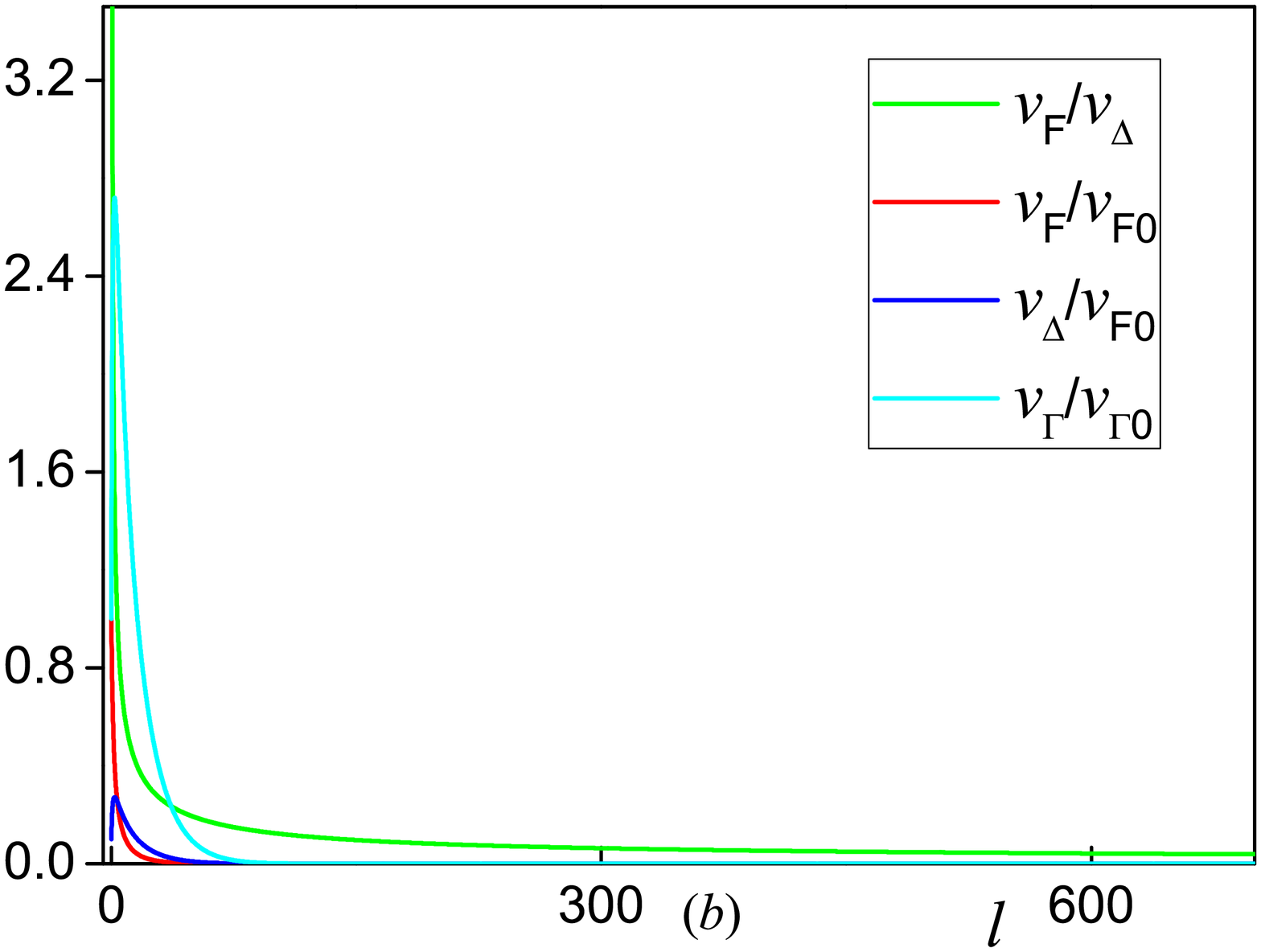,height = 6.15cm,width=9.2cm}\\ \vspace{-0.8cm}
\caption{(Color online) Flows of $v_F,v_\Delta,v_\Delta/v_F,$ and $v_\Gamma$ for the $\tau^x$
component in the presence of random gauge potential at two representative initial values $v_{\Delta0}/v_{F0}=0.1$ and $g=10^{-3}$ (The conclusions are independent of these concrete values) for: (a) $M=\tau^y$; (b) $M=\tau^z$. Their $\tau^z$-component counterparts are very similar to this case, and thus are not presented.} \label{Fig_gauge}
\end{figure}

By paralleling the clean limit, the coupled RG equations consisting of the fermion
velocities $v_F$, and $v_\Delta$, and disorder strength parameter $v_\Gamma$ derived
in Sec. \ref{Sec_RG-analysis} can be numerically solved. The disorder effects at a
nematic quantum critical point ($M=\tau^x$) has been studied in Ref. \cite{WLK}.
In the following, the corresponding effects for the cases $M=\tau^y$ and $M=\tau^z$
are addressed. In the present problem, the combining factor $g v_\Gamma^2$ represents the
disorder strength, not only $v_\Gamma$, and our analysis is valid for weak disorders with
small $g v_\Gamma^2$, and thus both $g$ and $v_\Gamma$ are assumed to be small in magnitude.
To obtain the compact plots, we have measured $v_\Gamma$ with $v_{\Gamma0}$ in Figs. (\ref{Fig_mass_y}),  (\ref{Fig_mass_z}) and (\ref{Fig_gauge}).

The random mass is considered first. The numerical calculation is proven to
be costly and the decreasing rate of  $v_\Gamma$ is small in the case of $M=\tau^y$ as
presented in Fig. (\ref{Fig_mass_y}). Although the flows are shown in finite $l$,
the tendency is straightforward. The approximately analytical discussion also supports
this, which includes $v_\Gamma(l)/v_{\Gamma_0}\approx0.191132 \exp\{-3.65314\times10^{-5}
(113726+l)\}$ ($l>600$) at a representative value $g=10^{-3}$. Therefore, $v_\Gamma$ flows
to 0 when $l$ approaches infinity.  As can be easily seen from Fig. (\ref{Fig_mass_z})
for the case $M=\tau^z$, $v_\Gamma$ decreases more rapidly. Despite flowing a little slower,
the fermion velocity ratio $v_F/v_\Delta$ eventually vanishes in the low energy limit.

We next discuss the case of the random gauge potential. By carrying out analogous steps,
we come to the similar conclusion that the random gauge potential cannot qualitatively
change the running behavior of $v_F$, and $v_\Delta$ for both $M=\tau^y$ and $M=\tau^z$,
which are depicted in Fig. (\ref{Fig_gauge}).

\begin{figure}[t]
\centering
\epsfig{file=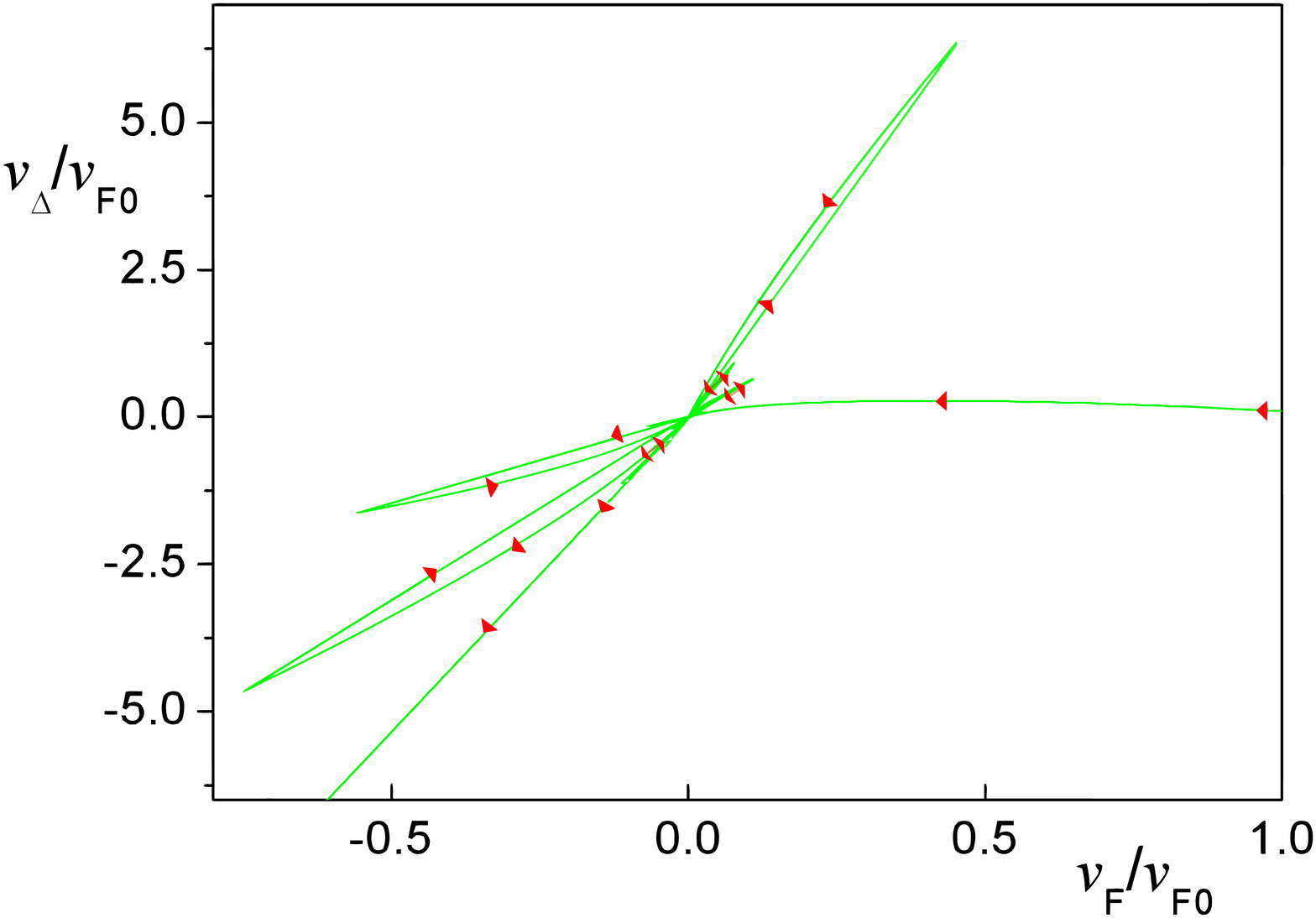,height = 6.15cm,width=9.2cm}\\ \vspace{-0.6cm}
\caption{(Color online) Flow of $(v_F,v_\Delta)$ for $l\in[0,600]$ in the presence of
random chemical potential at two representative initial values $v_{\Delta0}/v_{F0}=0.1$
and $g=10^{-3}$ (The conclusions are independent of these concrete
values). } \label{Fig_v_F-v_D} 
\end{figure}

Learning from Eq. (\ref{Eq_RG_vGamma}), we are informed that the disorder
strength parameter $v_\Gamma$, unlike the random mass and random gauge potential,
is marginal in the presence of a random chemical potential. (Since this property
is independent of the types of matrix states, the indices denoting types of matrix states
are discarded without loss of generality in the following). This indicates that $v_\Gamma$
does not flow as $l$ grows and hence should be kept as a constant. Consequently, the
influence of scattering due to the random chemical potential can not be neglected. As
expected, the flows of velocities $v_F$ and $v_\Delta$ are heavily dependent on
the magnitude of $v_\Gamma$. According to expression (\ref{Eq_RG_vDF}), it seems
that the running behavior of the velocity ratio $v_\Delta/v_F$ (or $v_F/v_\Delta$) is
independent of the disorder strength $v_\Gamma$. However, this is artificial. In the
present problem, the flow equation of $v_\Delta/v_F$ (or $v_F/v_\Delta$) is derived
from the more fundamental equations of $v_\Delta$ and $v_F$, and therefore, is reliable
only when $v_\Delta$ and $v_F$ both have well-defined fixed points \cite{Tesanovic,Chubukov}.
If the RG equations of $v_\Delta$ and $v_F$ have unphysical values, the running equation of $v_\Delta/v_F$ (or $v_F/v_\Delta$) becomes meaningless. Based on numerical
calculations in the presence of random chemical potential, the flow of $v_F$ and $v_\Delta$
in $v_F-v_\Delta$ space at two representative initial values is presented in Fig. (\ref{Fig_v_F-v_D}). This implies that the unphysical values of $v_F$ and $v_\Delta$
are generated with increasing $l$. Furthermore, they manifest rapid oscillations between
positive and unphysical negative values as $l$ grows. Therefore, $v_F$ and $v_\Delta$
do not reach any stable values due to the interaction between nodal
QPs and the random chemical potential.

In order to understand this concretely , we would also address a briefly
qualitative analysis. The flow equations of $v_F$ and $v_\Delta$ in the presence of
the random chemical potential are
\begin{eqnarray}
\frac{dv_{F}}{dl} &=& \left(C_{1}-C_{2}-\frac{v_{\Gamma}^{2} g}{2\pi v_{F}v_{\Delta}}\right)v_{F}, \\
\frac{dv_{\Delta}}{dl} &=& \left(C_{1}-C_{3}-\frac{v_{\Gamma}^{2} g}{2\pi v_{F}v_{\Delta}}\right)v_{\Delta},
\end{eqnarray}
where $g$ and $v_\Gamma$ correspond to the random chemical case. In the spirit
of RG analysis \cite{Shankar}, we can obtain the possible fixed pints of fermion
velocities $v_\Delta$ and $v_F$ by requiring that
\begin{eqnarray}
\frac{dv_F}{dl} &=& (C_1-C_2)v_F - \frac{v^2_\Gamma g}{2\pi
v_\Delta} = 0, \label{66} \\
\frac{dv_\Delta}{dl} &=& (C_1-C_3)v_\Delta - \frac{v^2_\Gamma
g}{2\pi v_F} = 0.\label{67}
\end{eqnarray}
We assume that $v_F^{*}$ and $v_\Delta^{*}$ correspond to the fixed
points. If both $v_F^{*}$ and $v_\Delta^{*}$ are finite, then the
above equations imply that $(C_1-C_2)v_\Delta^{*} v_F^{*} =
(C_1-C_3)v_\Delta^{*} v_F^{*}$, which can not be satisfied since
$v_\Delta^{*} \neq 0$. If $v_\Delta^{*} = 0$, then
\begin{eqnarray}
v_F^{*} = \frac{v^2_\Gamma g}{2\pi v_\Delta^{*}(C_1 - C_2)}.
\end{eqnarray}
From the expressions for $C_1$ and $C_2$, this implies that $1
\propto 1/(v_\Delta^{*})^2$, which is clearly inconsistent with the
assumption of $v_\Delta^{*} = 0$. Before going to the $v_F^{*} = 0$
case, we define $F_i=(v_F/v_\Delta)C_i,i=(1,2,3)$, then the new
forms of Eqs. (\ref{66}) and (\ref{67}) become
\begin{eqnarray}
(F_1-F_2)v_\Delta&=&\frac{v^2_\Gamma g}{2\pi v_\Delta} , \\
(F_1-F_3)\frac{v_\Delta^2}{v_F}&=&\frac{v^2_\Gamma g}{2\pi v_F} .
\end{eqnarray}
If $v_F^{*} = 0$, by both analytical and numerical analysis, we
found that these equations have no solution.

In conclusion, the fermion velocities $v_F$ and $v_\Delta$, as discussed above, do not
approach any stable values under the low energy regime caused by the interaction between
fermions and random chemical potential as shown in Fig. (\ref{Fig_v_F-v_D}). Therefore,
there is no fixed point of the fermion velocities $v_F$ and $v_\Delta$ in this case. We
interpret this as an indicator of the instability of the quantum phase transition in the
presence of the random chemical potential.

For completeness, we would like to make brief statements with various types of disorders
present. In general, there may be a certain number of kinds of disorders in realistic
physical problems \cite{Ludwig,Nersesyan,Stauber,Nayak,GZL,Khveshchenko}. First, we consider
a case in presence of the random chemical potential and random mass (and/or random gauge potential). Since the random mass and random gauge potential are both irrelevant, the marginal random chemical potential dominates and destroys the fixed points \cite{Stauber}. On the other hand, the corresponding fixed points would remain for a combination of two irrelevant cases of random mass and random gauge potential. Therefore, we can obtain the overall effects of disorders by investigating the three types of disorders separately.

\section{Velocity renormalization effects on superfluid density and critical temperature}\label{sec_quantities}

In an actual $d$-wave cuprate superconductor, the fermion velocities $v_F$ and $v_{\Delta}$
of the gapless nodal QPs are not equal. Indeed, the ratio $v_{\Delta}/v_{F}$
may be as low as $1/20$ \cite{Orenstein,Chiao}. As mentioned in Sec. \ref{Sec_numerical},
the value of this ratio $v_{\Delta}/v_{F}$ will be extremely influenced by $X$-ordering fluctuations in the proximity of quantum critical points. Since the ratio plays an important
role by entering a number of physical quantities \cite{Orenstein}, it is natural
to expect numerous intriguing effects due to the significant changes in the ratio $v_{\Delta}/v_{F}$. In this section, we primarily address the behavior of superfluid
density and critical temperature at different quantum critical points.

The superfluid density is an important quantity that characterizes
the fundamental feature of the $d$-wave superconducting state. In
cuprate superconductors, the superfluid density is known to exhibit
a linear $T$-dependence as \cite{Hardy,Lee97}
\begin{eqnarray}
\frac{\rho^s(T)}{m} = \frac{\rho^s(0)}{m}-\frac{2\ln2}{\pi}\frac{v_{F}}{v_{\Delta}}T,
\end{eqnarray}
with a coefficient proportional to the inverse of the velocity ratio $v_{\Delta}/v_{F}$.
Here, $\rho^s(0)=x/a^2$ is the superfluid density at zero temperature in the underdoped
region \cite{Orenstein,Orenstein90}, where $x$ and $a$ represent the doping concentration
and lattice spacing, respectively. For non-interacting nodal QPs, the velocity ratio
takes a bare constant, $v_{\Delta}/v_{F} \approx 0.1$ \cite{Orenstein,Chiao}, as
mentioned above.

An intimately related quantity is the critical temperature
$T_c$ \cite{Orenstein00PRL,Orenstein90,Hardy},
\begin{eqnarray}
T_c \propto \rho^s(0) \propto \frac{v_\Delta}{v_F} x,
\end{eqnarray}
where $x$ is the doping concentration.

Close to the critical point of the $X$ matrix state, due to
the strong $X$-order-parameter fluctuations, the velocity ratio
flows upon lowering the energy scale. The effects of velocity
renormalizations should be taken into account. To study these
in detail, after including the flow of velocities, Eqs.
(\ref{Eq_RG_vF}, \ref{Eq_RG_vD}, \ref{Eq_RG_vDF}), we can obtain
the renormalized superfluid density \cite{Lee97,Uemura,Liu},
\begin{eqnarray}
\rho^s_R(T)=\rho^s(0)-\rho^n_R(T),
\end{eqnarray}
\begin{eqnarray}
\frac{\rho^s_R(T)}{m}=\frac{4}{k_BT}\int \frac{d^2\mathbf{k}}{(2\pi)^2}
\frac{v_F^2(k)e^{\frac{\sqrt{v_F^2(k)k_x^2+v_\Delta^2(k)k_y^2}}{k_BT}}}
{\left(1+e^{\frac{\sqrt{v_F^2(k)k_x^2+v_\Delta^2(k)k_y^2}}{k_BT}}\right)^2},
\end{eqnarray}
where the velocities $v_{F,\Delta}$ are very complexly dependent on $k$ and
determined by the running equations in Sec. \ref{Sec_RG-analysis}.  The
renormalized critical temperature can be acquired by setting
\begin{eqnarray}
\rho^s(0)=\rho^n_R(T_c).
\end{eqnarray}
To estimate these more quantitatively, we assume the ultraviolet cutoff
$\Lambda= 10\mathrm{eV}$, and choose the representative bare velocity ratio
$v_\Delta/v_F \approx 0.1$, which is an appropriate value for
YBa$_2$Cu$_3$O$_{6+\delta}$ \cite{Chiao}.

\begin{figure}[t]
\centering
\epsfig{file=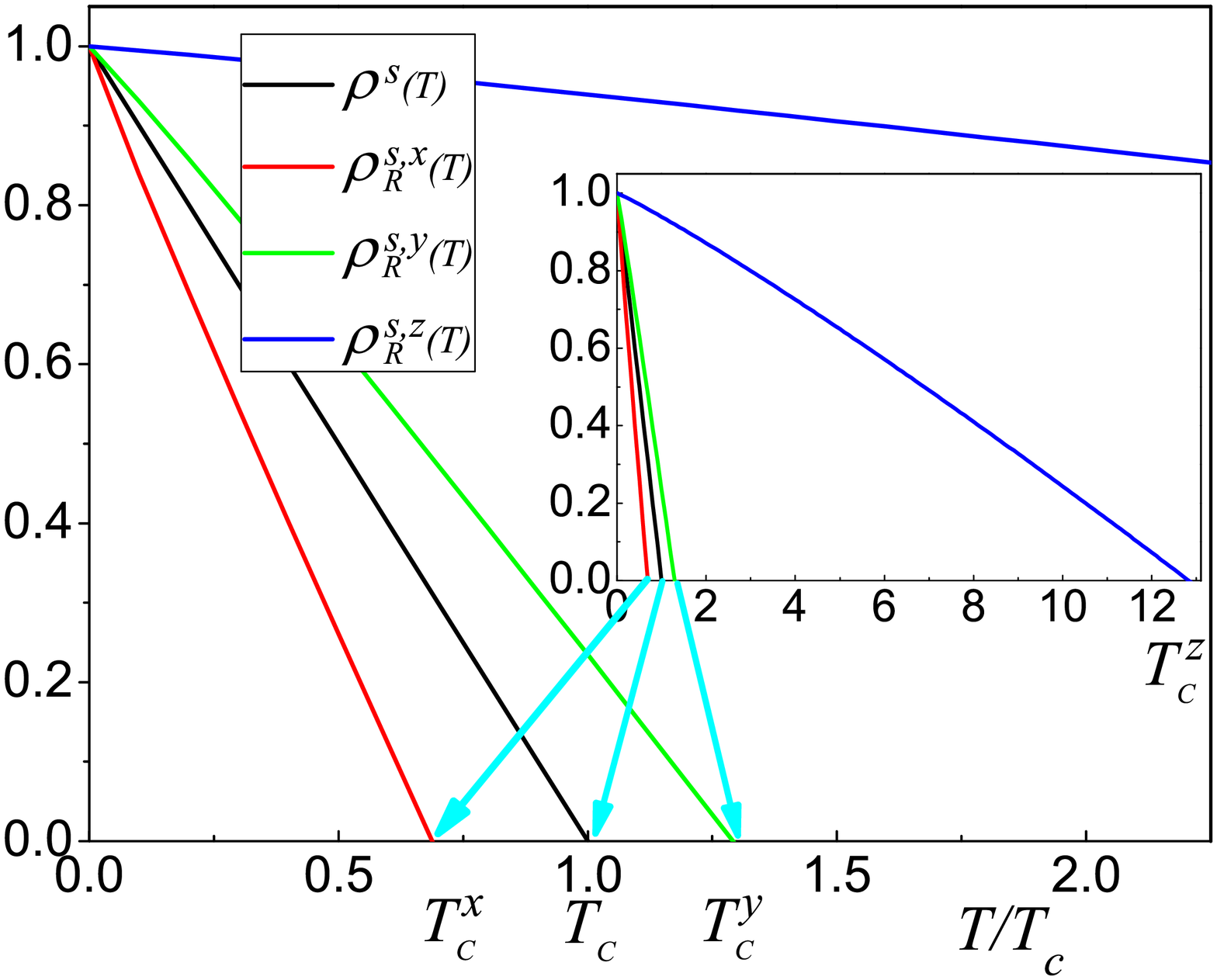,height = 6.25cm,width=9.5cm}
\vspace{-0.8cm}
\caption{(Color online) Superfluid densities [in units of zero-temperature
superfluid density $\rho^s(0)$] are calculated for $\Lambda/T_c=10^4$ and
$v_{\Delta0}/v_{F0}=0.1$. $\rho^s(T)$ denotes the superfluid density in
the absence of effects of quantum fluctuations. $\rho^{s,x}_R(T)$,
$\rho^{s,y}_R(T)$, and $\rho^{s,z}_R(T)$ correspond to the renormalized
superfluid density for $M=\tau^x$ (nematic), $M=\tau^y$, and $M=\tau^z$,
respectively. The critical temperatures labeled $T^{x,y,z}_c$ were obtained
directly from the intersections of the curves of superfluid densities with
the horizontal axis. It is obvious that both the superfluid density and the
critical temperature for $M=\tau^x$ are suppressed. However, their counterparts
for $M=\tau^{y}$ receive slight enhancements, and those for $M=\tau^{z}$,
significant enhancements.}\label{Fig_tem-rho}
\end{figure}

First, we consider the clean limit system. Approaching the critical point of the
$X$ matrix state, the velocity ratio flows to $v_{\Delta}/v_{F}=0$ \cite{Huh},
$v_{\Delta}/v_{F}=1$, and $v_{F}/v_{\Delta}=0$, corresponding to $M=\tau^x$, $\tau^y$,
and $\tau^z$, respectively. This leads to the suppression of the superfluid density
in case the $M=\tau^x$ \cite{Liu} and growth in the rest of the cases. The numerical results
for superfluid densities and critical temperatures are shown in Fig. (\ref{Fig_tem-rho}).
As mentioned, based on the fixed point $v_\Delta/v_F\rightarrow0$ for $M=\tau^{x}$, the superfluid density and critical temperature are both suppressed. On the other hand, they
are slightly increased in the case of $M=\tau^y$. Furthermore, to the case $M=\tau^z$, it
exhibits substantial enhancements owing to the running flow to $v_F/v_\Delta\rightarrow0$.
Since plenty of approximations are unavoidably employed during the calculations, we need
to point out here that the quantitative enhancements may not be reliable. However, the
qualitative increments of superfluid density and critical temperature would be unambiguous.

Next, we discuss the effects caused by disorders, including random mass, random
gauge potential, and random chemical potential. According to the analysis in Sec. \ref{Subsec_disorder}, the disorder strength $v_\Gamma$ is irrelevant both
for random mass and for random gauge potential. Therefore, the behavior of the superfluid
density and critical temperature mentioned in the previous paragraph are unchanged in
the presence of these two types of disorders. By comparison, quantum phase transitions
are unstable against the random chemical potential. This suggests that the corresponding
fixed points do not exist.

In brief, we conclude that distinct behaviors of physical quantities, such as
superfluid density and critical temperature, will be generated by the influence
of singular velocity renormalization near the presumable $X$-ordering quantum
critical points. As a consequence, this leads to a possible way to determine the
$X$-ordering quantum critical points by seeking these singular behaviors. This
may be of great help to understand the complicated phase diagram and many anomalous
features of high-temperature superconductors.

\section{Summary}\label{Sec_summary}

In summary, we investigate unusual renormalizations of the Fermi velocity $v_F$ and
gap velocity $v_{\Delta}$ of nodal QPs close to diverse quantum critical points
classified by vertex matrices (\ref{Eq_S_psi-phi}) in the $d$-wave superconducting
dome. According to the spirit of the RG method \cite{Huh,Shankar,WLK}, the series of RG
equations$\raisebox{0.1mm}{---}$both in clean limit and in the presence of three types
of static disorders, namely, random mass, random gauge potential, and random chemical$\raisebox{0.1mm}{---}$are derived to one-loop level. A detailed RG
analysis and numerical computations are given.

In clean-limit systems, these RG equations lead to three distinct fixed points of
the velocity ratio: $v_\Delta/v_F\rightarrow0$ \cite{Huh}, $v_\Delta/v_F
\rightarrow1$, and $v_\Delta/v_F\rightarrow\infty$ for $M=\tau^x$, $\tau^y$, and
$\tau^z$, respectively. Since the velocity ratio, $v_\Delta/v_F$, enters various
physical quantities, these fixed points result in abundantly physical observable
effects. In particular, we calculate the superfluid density and critical temperature.
They are affected sensitively by approaching these fixed points. Superfluid densities
and critical temperatures are both suppressed, slightly increased, and significantly
enhanced for cases $M=\tau^x$, $\tau^y$, and $\tau^z$, respectively.

Three types of disorder effects on these fixed points are also examined. We present
a series of coupled RG equations of Fermi velocity $v_F$, gap velocity $v_\Delta$,
and disorder strength parameter $v_\Gamma$. After both analytical and numerical
computations, we find that fixed points obtained in the clean limit are robustly stable
in the presence of random mass and random gauge potential. However, these fixed points
are destroyed in the presence of a random chemical potential which is marginal. Therefore,
this is responsible for the instabilities of quantum phase transitions. The corresponding
effects on physical quantities are also studied. Compared to the clean limit, the behaviors of physical quantities controlled by fixed points, such as the superfluid density and critical temperature depicted in Fig. (\ref{Fig_tem-rho}), are unchanged when a random mass or random gauge potential is included. On the other hand, the effects of quantum critical fluctuations on physical observables can be neglected due to the fixed points broken
in the presence of a random chemical potential.

The occurrence of quantum phase transition stems from competition among ground state phases \cite{Vojta2003RPP,Sachdev2011PT,Sachdev2011book}. The physical behavior of a system is influenced significantly within a wide scope of the phase diagram, especially near quantum critical points where fluctuations are divergent \cite{Vojta2003RPP}. By studying various
quantum phase transitions in a superconducting dome, the critical behavior of physical quantities can be captured. Remarkably, this provides a helpful clue to confirm or even locate the very existence of quantum critical points by means of detecting these singular behaviors. Therefore, it will be instructive to understand the complicated phase diagram and a number of anomalous features of high-temperature superconductors.

\section*{ACKNOWLEDGEMENTS}

I am very grateful to Guo-Zhu Liu for his supervision and encouragement. In addition, the author acknowledges the MPI for Solid State Research for cordial hospitality during his stay. This work was supported by the National Natural Science Foundation of China under Grants Nos. 11074234, and 11274286, and the joint doctoral promotion program sponsored by the Max Planck Society and the Chinese Academy of Sciences.


\end{document}